\documentclass{amsart}
\newtheorem{theorem}{Theorem}[section]

\newtheorem{corollary}[theorem]{Corollary}
\newtheorem{proposition}[theorem]{Proposition}

\theoremstyle{definition}
\newtheorem{definition}[theorem]{Definition}
\newtheorem{example}[theorem]{Example}

\theoremstyle{remark}

\newcommand{\cA}{{\mathcal A}}
\newcommand{\cB}{{\mathcal B}}
\newcommand{\C}{{\mathcal C}}
\newcommand{\cD}{{\mathcal D}}
\newcommand{\cL}{{\mathcal L}}

\newcommand{\cF}{{\mathcal F}}
\newcommand{\cH}{{\mathcal H}}

\newcommand{\cK}{{\mathcal K}}
\newcommand{\cM}{{\mathcal M}}

\newcommand{\cP}{{\mathcal P}}

\newcommand{\cR}{{\mathcal R}}
\newcommand{\cS}{{\mathcal S}}

\newcommand{\cW}{{\mathcal W}}
\newcommand{\cV}{{\mathcal V}}
\newcommand{\cZ}{{\mathcal Z}}
\newcommand{\cJ}{{\mathfrak{J}}}

\newcommand{\bR}{{\mathbb{R}}}
\newcommand{\bC}{{\mathbb{C}}}
\newsymbol\boxtimes 1202
\newcommand{\Cs}{{{$\hbox{\bf C}^*$}}}
\newcommand{\Fs}{{F_{max}(p_{\xi},\eta)}}
\numberwithin{equation}{section}

\newcommand{\Tr}{\mathrm{Tr}}
\newcommand{\id}{\mathrm{id}}
\newcommand{\vr}{\varrho}

\newcommand{\jed}{{\mathbb{I}}}
\begin{document}

\setlength{\baselineskip}{2\baselineskip}

\begin{titlepage}
\begin{center}
\vspace*{15mm}
{\LARGE POSITIVE MAPS, STATES, ENTANGLEMENT AND ALL THAT;\\
 {\it some old and new problems}.
}\\
\vspace{2cm}
\textsc{{\large W{\l}adys{\l}aw A. Majewski}\\
Institute of Theoretical Physics and Astrophysics\\
Gda{\'n}sk University\\
Wita Stwosza~57\\
80-952 Gda{\'n}sk, Poland}\\
\textit{E-mail address:} \texttt{fizwam@univ.gda.pl}\\
\end{center}
\vspace*{1cm}
\textsc{Abstract.}
We  outline a new approach to the characterization
as well as to the classification of positive maps.
This approach is based on the facial structures of the set of states and 
of the cone of positive maps. In particular, the equivalence between Schroedinger's and Heisenberg's
pictures is reviewed in this more general setting. Furthermore, we discuss in detail the structure
of positive maps for two and three dimensional systems. In particular, the explicit form of decomposition
of a positive map and the uniqueness of this decomposition for extremal positive maps 
for 2 dimensional case are described. The difference of the structure of positive maps 
between 2 dimensional and 3 dimensional cases is clarified.  
The resulting characterization of positive maps is
applied to the study of quantum correlations and entanglement.

\vspace{1cm} {\bf Mathematical Subject Classification}: Primary: 46L53, 46L60:
Secondary: 46L45, 46L30

\vspace{0.5cm}
 
\end{titlepage}

\newpage
\section{INTRODUCTION}
The aim of this paper is to bring together two areas, the theory of positive maps on 
\Cs-algebras and the abstract characterization of the set of states on a \Cs-algebra.
The present paper concerns crucial aspects of the quantization procedure;
as such it is an extension of our recent publication \cite{Open}.
The classification of positive maps and the full characterization of 
states are at the heart of quantum theory, with particular reference to
the foundations of quantum information theory \cite{Albert}, \cite{Keyl}. 
To be more precise, the full description 
of the set of states of a physical system, the complete characterization
of distinct types of states, and the detailed account of the properties of maps of states into states
(i.e. the Schroedinger approach to dynamical maps) involve various aspects
of the affine structure of the convex set of (all) states. 
The standard tool for a study of convex compact sets is Krein-Milman theorem which states that 
such a set is the closure of  the convex hull of its extreme points. 
In particular, this idea was used by St$\o$rmer \cite{St} for a classification of positive maps.
But, such an approach depends on the description of extreme positive maps. The characterization 
of extreme positive maps is available 
only for the $M_2(\bC)$-case ($M_n(\bC)$
stands for the set of all $n$ by $n$ matrices with complex entries).  
Hence it is natural to go one step further 
and consider the set of positive maps or of states
as a (convex compact) subset of an ordered Banach space.
The characterization of certain subsets of the set of all states would provide 
a nice illustration of such method.
In particular, using the Krein approach to the geometric version of Hahn-Banach theorem, one can 
introduce special functionals as basic tools for a characterization of some subsets of states.
As instance of this is witnesses of entanglement (cf. page 452 in \cite{Keyl}) 
to study entanglement of states.

Further, we can 
combine ordered Banach space techniques (defined for the state space) with the theory of
linear positive unital maps and exploit
the algebraic structure of the underlying algebra.
In this manner we can get a better understanding of the structure of subsets of states. 
This is due to the fact, that the ``plain'' language of ordered Banach spaces, in general, does not  ``feel'' 
the non-commutativity 
of the underlying algebra. Consequently, there is a need for a supplementary geometrical structure 
of the set of states to establish rigorous relations between the theory given in terms of states and
the algebraic structure of the set of observables, i.e. to get a more complete understanding of the nature of
``the equivalence'' between Schroedinger and Heisenberg's pictures.
We note that the need for a clarification of such an equivalence
is not new; 
Dirac \cite{PAMD} working within the context of quantum electrodynamics had already noted some problems 
connected with it. As another example,
we wish to point out problems emerging from the description of quantum chaos, \cite{chaos}.
In this paper we will provide the mathematical argument showing the necessity for
a new look upon the discussed equivalence (see Section 3).

Our approach will stem from the so called Kadison question: {\it under what conditions a convex set is 
affinely isomorphic to the set of states on a Jordan {\rm(} \Cs, $W^*$ respectively {\rm )} algebra?}
(see the Glossary in the Appendix). We recall that the abstract characterization of the set of observables
based on Jordan algebras
(Heisenberg picture) was established by von Neumann, Jordan and Wigner
 some seventy years ago (see \cite{JNW} and \cite{Neumann}) 
while the essence of the abstract
characterization of the set of states (Schroedinger picture) is contained in the Kadison question.
The full answer to the Kadison question, provided recently by Alfsen and Shultz in
\cite{A} and \cite{A2}, will be our starting point.

\smallskip

We will show
that pure quantum features of non-commutative dynamical systems such as the peculiar behaviour of positive maps,
quantum correlations and entanglement can be more easily understood
within the mathematical framework which will be introduced in the subsequent sections. 
The main idea of our approach to the description of distinct types of states as well as to 
that of positive maps 
is to replace small boundary subsets (extreme points) by
larger subsets for which we have an explicit description.
In particular, we propose a modification of St$\o$rmer's approach to the
classification of positive maps; namely to replace extreme positive maps by maximal
faces of (n-) positive maps; for the appropriate definitions see Section 2.
Here, we only note that an extreme point is a face, thus it is contained in a maximal face.
As we will see in Section 4, Kye gave \cite{K1} - \cite{Kyecan} the complete characterization of 
appropriate (i.e. maximal) faces. This points to the choice of maximal faces as a powerful tool 
for our purpose.

The paper is organized as follows.
In Section 2 we review some of the standard facts on theory of convex sets
and set up notation and terminology. Our presentation is entirely based on 
 two fundamental books by Alfsen-Shultz \cite{A} and \cite{A2}.
Section 3 is devoted to the study of positive maps from the physical point of view.
Again, using Alfsen-Shultz monographs we will compare Schroedinger's  and Heisenberg's picture to 
show that there is a one-to-one
correspondence in the description of decomposable maps (in both) pictures if and only
if one equips the Schroedinger picture with additional geometrical structure.
The relations between the facial structures of states and positive maps
are given in Section 4. Section 5 presents basic properties of positive maps while Section 6
concerns low dimensional cases. 
This section is based on a joint work with Marcin Marciniak and its aim is to get 
a deeper understanding of the reason why
the theory of positive maps changes so dramatically when one goes
from 2-level systems to 3-level ones.
We present the explicit form of decomposition
of a positive map and the uniqueness of this decomposition for extremal positive maps,  
for 2 dimensional case, are described. Furthermore, the difference of the structure of positive maps 
between 2 dimensional and 3 dimensional cases is clarified from the geometrical point of view.  
The last section contains a brief discussion of applications of our results
to the description of entangled states of quantum systems.
It is worth pointing out that our approach sheds new light on the construction of
non-decomposable maps which is an important issue in any attempt to  classifying
entangled states. 

Finally, we want to stress that, in order to make the paper more accesible to a quantum computing audience,
 we shall
deliberately not address the problem in its full generality. Consequently, although the theory may be formulated 
in general \Cs-algebraic terms,  we will be interested mainly in $\cB(\cH)$ (i.e. in the \Cs-algebra 
of all linear bounded operators on a Hilbert space $\cH$).

\section{GEOMETRY OF STATE SPACES}
Let $F$ be a convex subset of a convex set $\cS$ in some Banach space. 
$F$ is said to be a face of $\cS$ if the following 
property holds:
\begin{equation}
x,y \in \cS, \, (1-t)x + ty \in F \quad for \; some \;\; t \in (0,1)\quad \Longrightarrow \quad x,y \in F.
\end{equation}

A proper face $F$ is a face of $\cS$ which is neither $\cS$ itself nor the empty set.
Note that a face of a face of a convex set $\cS$ is a face. It is also clear that the intersection 
of faces is again a face. Therefore, there is a unique smallest face contained in a given subset.
Also, given a family $\{F_i; i \in I \}$ of faces, we denote by $\vee_{i \in I}F_i$, the smallest face 
containing every $F_i$. Hence, the set $\cF(\cS)$ of all faces of a convex set $\cS$ is a complete 
lattice (see A17 \footnote{In the following, reference marks like this in the text refer to 
definitions and basic facts listed in the Glossary at the Appendix given at the end of the paper.}) 
 with respect to the partial order induced by the set of inclusions.

\smallskip

Here and below, $\cS$ will denote the set of all states on a \Cs-algebra $\cA=\cB(\cH)$, 
namely the convex set of all normalized, positive linear maps $\vr : \cA \mapsto \bC$, on $\cA$.
$face(\vr)$ will stand for the face generated by the state $\vr$, i.e. the smallest non-trivial 
convex set of convex decompositions of $\vr$, $\vr = \sum_i \lambda_i \sigma_i$, $\lambda_i \ge 0$, 
$\sum_i \lambda_i = 1$, into other states $\sigma_i$.  
Let $(\cH_{\vr}, \pi_{\vr}, \Omega_{\vr})$ be the GNS triple (see A9) associated with a 
state $\vr$ on a \Cs-algebra $\cA$. Then, one has the following nice characterization
of the face $face(\vr)$. Namely, for every positive functional
$\sigma \in face(\vr)$ there exists a unique positive element $b \in \pi_{\vr}(\cA)^{\prime}$
such that 
\begin{equation}
\sigma(a) = (\Omega_{\vr}, b \pi_{\vr}(a) \Omega_{\vr}) \quad for \quad all \quad a \in \cA.
\end{equation}
Here, $\pi_{\vr}(\cA)^{\prime}$ stands for the commutant of $\pi_{\vr}(\cA)$; (see A4).
Moreover, the map $\phi: \sigma \mapsto b$ is an order preserving affine isomorphism (see A1) of
 $face(\vr)$ onto $(\pi_{\vr}(\cA)^{\prime})_+ (\equiv \{ a \in \pi_{\vr}(\cA)^{\prime}, a \ge 0 \})$, 
i.e. $\phi$ is the affine isomorphism such that 
$\sigma \le \sigma^{\prime}$ implies $\phi(\sigma)= b \le b^{\prime} = \phi(\sigma^{\prime}).$

To proceed with the discussion of the geometry of state space we will need two 
concepts. The first one is the so called 
projective face which can be characterized as follows. Let $F$ be a norm closed
face in $\cS$. If $p$ is the carrier projection of $F$ (the smallest projection $p$ such that
$\sigma(p) = ||\sigma||$ for all $\sigma \in F$)
then $F \equiv F_p$ where
\begin{equation}
 F_p = \{\sigma \in \cS; \sigma(p) = 1 \}.
\end{equation}
A face of the form $F_p$, where $p$ is a projection in $\cA$, will be called a projective face.
This concept {\it can be defined in much more general setting}, namely for a pair of ordered
unit space and base norm space, for details see \cite{A2} and A16 for the terminology.

Let $p \in \cB(\cH)$ be an orthogonal projection. Then the map $ p \mapsto F_p$ 
determines an isomorphism from the lattice of closed subspaces of $\cH$ to the lattice of norm 
closed (projective) faces of $\cS$. The closed faces associated with a projection have 
another interesting property which will be useful to fully understand the 
Alfsen-Shultz result. Namely, if $p$ is a projection
onto the closed subspace spanned by a family of unit vectors $\{ \eta_i \}_{i \in I} \subset \cH$,
then the norm closed face $F$ associated with $p$ is the smallest face of $\cS$ that
contains the vector states $(\omega_i )_{i \in I}$  where 
$\omega_i(\cdot) = (\eta_i, \; \cdot \; \eta_i)$.

As a corollary one has the following result. 
The face generated by two distinct pure states
of the normal state space of $\cB(\cH)$ is an Euclidean 3-ball, i.e. the face is affinely
isomorphic to the closed unit ball in Euclidean 3-dimensional space.

\smallskip

The second concept, {\it orientation}, 
is another important ingredient of the affine structure of the set $\cS$. 
For the sake of conciseness, we present this concept only briefly, as
it is a necessary "tool" for understanding the relations
between the affine structure of the state space $\cS$ and the Jordan 
and Lie products of the corresponding
algebra. 

To illustrate this idea let us consider the algebra of all $2$ by $2$ matrices with 
complex entries, $M_2(\bC)$, so a very special example of $\cB(\cH)$.
It can be shown \cite{A} that
the affine structure of the state space determines the Jordan product on $M_2(\bC)$
uniquely. There are two possible \Cs-products, both being well
defined Jordan products: the usual one $M_2(\bC) \times M_2(\bC) \ni <a,b> 
\mapsto ab \in M_2(\bC)$ and the opposite one  $M_2(\bC) \times M_2(\bC) \ni <a,b>
\mapsto ba \in M_2(\bC)$. 
However, the definition of positive elements in $M_2(\bC)$ is {\it not affected by the above
ambiguity}. This clearly shows that an additional concept is necessary to determine the form of the 
associative product in the algebra. 

\smallskip

Let us be more formal and provide some further tools necessary to the 
geometrical characterization of the state space.
 A self-adjoint operator 
$s \in \cB(\cH)$ is called (e-) symmetry if $s^2 = 1$ ($s^2 = e$, $ e$ a projector, 
respectively). Then, to each symmetry with canonical 
(spectral) decomposition $s = p - q$ ($p,q$ are orthogonal projectors,  $pq=0$, and
$p+q =1$ or $p +q = e$ respectively) we assign 
the pair of projective faces $(F_p, F_q)$ called the associated
generalized axis. Having the concept of (e-) symmetries one can generalize the idea of orthogonal 
frame of axes in the state space $\cS(M_2(\bC))$ which, we recall, is affine isomorphic to
the ball in 3 dimensional Euclidean space.
In general, a triple of symmetries $(r,s,t)$ is called a Cartesian triple \footnote{If $r,s,t$ are 
e-symmetries then a Cartesian triple of e-symmetries is a Cartesian triple of symmetries
in the von Neumann algebra $e \cB(\cH) e$.} if the following
conditions are satisfied:
\begin{itemize}
\item $r \circ s = s \circ t = t \circ r = 0$, where $\circ$ stands for the Jordan product.
\item $U_rU_sU_t = id$, $id$ stands for the identity operator while $U_v a \equiv vav$
for any $a \in \cB(\cH)$ and any symmetry $v \in \{r,s,t \}$.
\end{itemize}

\smallskip

A nice example of a Cartesian triple of symmetries for $M_2(\bC)$ 
is provided by Pauli spin matrices.
The question of existence of Cartesian triples is settled by the following result: a von Neumann 
algebra $\cM$ (so in particular, $\cB(\cH)$) contains a Cartesian triple 
of e-symmetries if and only if $e$ is a halvable projector, i.e. $e$ is a sum of two equivalent 
(in the von Neumann sense, see A13) projectors.
Then, $F_e$ is affinely isomorphic to the normal space of the local algebra $e\cM e$.

\begin{example}
$M_4(\bC)$ contains two families of Cartesian triples of e-symmetries. One for $e$ of rank 
2 and the one for identity. Note, that for $M_2(\bC)$ one has only one family of discussed 
triples: the one which is associated with $\jed$.
\end{example}

Now we can define the previously mentioned concept of orientation for a von Neumann algebra $\cB(\cH)$.
First, the local orientation of $\cB(\cH)$ is a unitary equivalence class of Cartesian
triples in $e \cB(\cH) e$ where $e$ is a halvable projection in $\cB(\cH)$.
Then the global orientation is defined as a "continuous choice" of local orientations.
It can be proved (cf. \cite{A}) that {\it there is one-to-one correspondence between global orientations 
of $\cB(\cH)$ and Jordan compatible associative products in $\cB(\cH)$}, i.e. the Jordan product
associated with the geometry of states coincides with the original product of $\cB(\cH).$

\smallskip

After these preliminaries we are in position to give the answer to Kadison question and to discuss, 
in the next Section, the equivalence between Heisenberg and Schroedinger's pictures.
Here, we will do it for $\cB(\cH)$ only (for a general treatment see \cite{A2}; 
see also A16, A10, A11, A15 for the terminology).

\begin{theorem} [Alfsen, Shultz, \cite{A2}]
\label{AlfsenShultz}
Let $K$ be the base of a complete base norm space. Then $K$ is affine isomorphic to
the normal state space, $\cS_0$, of $\cB(\cH)$ with $\cH$ a complex Hilbert space
if and only if the following conditions hold:
\begin{itemize}
\item every norm exposed face is projective;
\item the $\sigma$-convex hull of extreme points of $K$ equals $K$;
\item the face generated by every pair of extreme points of $K$ is a 3-ball and is norm exposed.
\end{itemize}
\end{theorem}

It is worth pointing out that since the $3$-ball constitutes the so called
Bloch sphere which coincides with the state space for the standard two-level system
{\bf the last condition of Theorem \ref{AlfsenShultz} clearly indicates
the fundamental role of qubits}. In other words, the set of ``two dimensional states'' 
plays ``locally'' a crucial role
in the general characterization of the set of all normal states over $\cB(\cH)$.
However, it should be stressed that for a general \Cs-algebra the face generated by a pair of 
pure states is either a 3-ball or a line segment. Thus, the above simple picture for 
$\cB(\cH)$ turns to be more complicated for a general \Cs-algebra.

\section{POSITIVE MAPS AND THEIR DUALS.}
Let $\cP_0$ denote the convex set of all $\sigma$-weakly continuous unital
positive linear maps (so such maps $\alpha$ that the state $\varphi \circ \alpha$ 
is determined by a density matrix whenever $\varphi$ has this property) from the von Neumann algebra 
$\cB(\cH)$ into itself; the subscript ``$_0$'' stands for unital.
We emphasize that contrary to the standard conjecture saying
that only completely positive maps have a direct interpretation as dynamical maps,
it seems that some maps in the 
class $\cP_0$ of plain positive maps could also be relevant for 
description of time evolution (see \cite{sudershan} for a recent discussion of this question; 
a recent survey on dynamics of open quantum systems can be found in the lecture notes \cite{Kup}).
Moreover, $\cP_0$ contains large subsets of positive maps which are directly connected with
a characterization of various types of entangled states what provides the additional motivation
for our interest in this class.

Let us turn to the question of dual (transposed) maps, i.e. maps
defined on the set of states.
Suppose $T \in \cP_0$ and 
define $(T^*\omega)(a) = \omega(Ta)$ where $a \in \cB(\cH)$ and $\omega$ is a normal
state on $\cB(\cH)$. Then

\begin{theorem}
There is a one-to-one correspondence between $\sigma$-weakly continuous
positive unital linear maps from $\cB(\cH)$ into itself, and affine maps from the normal 
state space of $\cB(\cH)$ into itself.
\end{theorem}

We now denote by $\cD_0$ the more specialized family, $\cD_0 \subset \cP_0$, 
of positive maps consisting of the so called decomposable maps.
The general form of such maps, in the Heisenberg picture is defined by the following relations

\begin{equation}
\label{3.1}
T(a) = \sum_i W_i^* \tau_i(a) W_i \quad(\equiv T_{\cW}(a)),
\end{equation}

where $W_i \in \cB(\cH)$, $\sum_i W_i^* W_i = \jed$, while $\tau_i$ stands for a unital Jordan homomorphism, i.e. 
$\tau_i$ is a linear map preserving the Jordan structure $\tau_i (\{a,b\}) = \{ \tau_i(a),
\tau_i(b) \}$, with $\{\cdot, \cdot \}$ standing for the anticommutator. In (\ref{3.1}), $\cW$ denote the 
subspace of $\cB(\cH)$ spanned by
$\{ W_1,...,W_n \}$.
The special case when the $\tau_i$'s  are $^*$-morphisms leads to the important class of
{\it completely positive maps}, $\C \cP_0$.
To pass to the dual picture (so to go to Schroedinger's picture), we need (see also A18, A17)

\begin{theorem}[Alfsen, Shultz \cite{Shultz}, \cite{A2}]
\label{ASJor}
Consider $\cB(\cH_1)$, $\cB(\cH_2)$ with normal state spaces $\cS_1$ and
$\cS_2$ respectively and let $T^*_0: \cS_2 \to \cS_1$ be an affine map. Let $T: \cB(\cH_1) \to \cB(\cH_2)$
be the unital positive $\sigma$-weakly continuous map such that $T^*|_{\cS_2} = T^*_0$ where
$T^*$ is defined by the formula:  $(T^*\omega)(a)
\equiv \omega(T(a))$ for any $a \in \cB(\cH_1)$ where $\omega$ is any linear normal functional 
on $\cB(\cH_2)$.
Then, the following statements are equivalent:
\begin{itemize}
\item  $T$ is a unital Jordan homomorphism from $\cB(\cH_1)$ into $\cB(\cH_2)$.
\item $(T^*_0)^{-1}$ preserves complements of projective faces.
\item $(T^*_0)^{-1}$ as a map from the lattice of projective faces of $\cS_1$ into the lattice 
of projective faces of $\cS_2$ preserves lattice operations and complements.
\end{itemize}
\end{theorem}

Consequently one has

\begin{corollary}
\label{wniosek1}
There is a one-to-one correspondence between the set $\cD_0$ of decomposable maps and 
the set of affine maps of the form
\begin{equation}
\omega \mapsto \sum_i U^*_{W_i}(T^*_{0,i}\omega) \bigl( \equiv
\sum_i \omega(W^*_i T(\cdot)W_i) \bigr)   
\end{equation}
where $U^*_{W_i}\omega(\cdot) = \omega(W^*_i \cdot W_i)$, $\omega$ any normal state, 
and $T^*_{0,i}$ satisfies one of the conditions given in Theorem \ref{ASJor}.
\end{corollary}

Finally, we want to describe the most "regular" case - the case of invertible maps; 
note that the hamiltonian time evolution is the best known example of such maps.

\begin{theorem}[Kadison \cite{Kad}]
\label{hamil}
Let $T^*$ be an affine invertible map from the state space $\cS$ of a \Cs-algebra $\cA$ 
(so also $\cB(\cH)$) onto itself. It follows that
there exists a unique Jordan automorphism $T$ of $\cA$ such that
\begin{equation}
(T^* \omega)(a) = \omega(Ta)
\end{equation}
for all $\omega \in \cS$ and $a \in \cA$.
\end{theorem}

Recall that any Jordan isomorphism can be splited into the sum of $^*$-isomorphism and $^*$-anti-isomorphism.
However, there is a possibility to distinguish between $^*$-isomorphism and $^*$-anti-isomorphism
on the Schroedinger picture level. Namely, employing the geometrical structure introduced in Section 2,
one has:

\begin{proposition}(Alfsen, Shultz \cite{A2})
Let $\Phi: \cB(\cH_1) \to \cB(\cH_2)$
be a Jordan isomorphism. Then $\Phi$ is a $^*$-isomorphism if and only if it preserves
orientation, and $\Phi$ is a $^*$-anti-isomorphism if and only if it reverses orientation.
\end{proposition}

\smallskip

Summarizing this section one has:

\begin{enumerate}
\label{sciany}
\item Plain positive maps as well as invertible dynamical maps are not sufficiently sensitive to 
the facial structure of states when one passes 
from Heisenberg's picture to Schroedinger's. This follows from the fact that for plain positive maps,
the ordered Banach space framework was used while for invertible maps, in Kadison theorem, the Jordan 
structure was indispensable ingredient.
\item On the other hand, for  decomposable positive maps the facial structure is essential.
\item $^*$-morphisms and $^*$-anti-morphisms can be distinguished on the set of states; in this case 
the geometrical 
structure of states plays again crucial role.
\item Let the time evolution be given in terms of a group. Then, the continuity properties of the group 
strengthen the conclusions stemming from  Kadison's result for 
hamiltonian type dynamics (cf. Theorem \ref{hamil}). Namely, one parameter group of affine maps on $\cS$
with suitably strong continuity properties gives rise to a group of $^*$-automorphisms and not merely 
to Jordan automorphisms (see \cite{BR}).
\end{enumerate}

Consequently, to guarantee the equivalence of the description of positive maps in both pictures
the Schroedinger picture 
should be equipped with the additional geometrical structure described in Section 2. 
{\bf This conclusion is all the more interesting in view of the fact 
that quantum computing needs decomposable maps.}
In particular, a description of entangled states may appeal to the specific geometrical features 
of the set of states. Finally, to comment (3) and (4) we note that to have the equivalence between
Schroedinger's and Heisenberg's pictures one should be able to determine the associative product on the
set observables not merely the Jordan structure. But, the observables in quantum mechanics are (quantum) 
random variables with a specified probability distribution for each state.
However, to determine evolution of observables 
(so to define non-commutative derivations as for example in the 
Heisenberg equation) one needs the Lie product. 
On the other hand, the Lie product with the Jordan product determine the associative product of 
the algebra of observables.
This clarifies the role of orientations
(cf also Proposition 3.5). As in this paper, the evolution of quantum systems will be not studied we skip 
the details.

\section{FACIAL STRUCTURES FOR STATES AND POSITIVE MAPS.} 
Having noted that the facial structure plays an essential role in the characterization of the set 
of all states, we turn to discussing
the facial structures of positive maps and their relations to the corresponding structures of states.
Throughout this Section we assume finite dimensionality of $\cH$ 
and consider $\cB(\cH)$, i.e. $M_n(\bC)$ where
$n = dim \cH$. First, we wish to indicate relations between projections in the set of
observables, so projections in $M_n(\bC)$ for the considered case, and faces of positive 
and completely positive maps,
so faces in $\cP$ and $\C\cP$, i.e. throughout this section we are going to consider 
positive maps which are not necessary unital.
The relations between facial structures of positive maps and of states are expected due to 
Theorem \ref{AlfsenShultz} and Corollary  \ref{wniosek1}. We begin with 
the following result:

\begin{theorem}[Kye \cite{Kyecan}]
Denote by $\cV$ the complete lattice of all subspaces of the $n$-dimensional vector 
space $\bC^n$, and by $\cJ(\cV)$ the complete lattice of all homomorphisms from $\cV$ into itself.
Then, there is a well-defined homomorphism 
$$\phi : \cF(\cP) \to \cJ(\cV)$$
where $\cF(\cP)$ is the complete lattice
of all faces of $\cP$.
\end{theorem}

To give a more specialized result we need the concept of matricially convex faces in
$\C \cP$. Let $T_i \in \C \cP$ and $ b_i \in M_n(\bC)$ for $1 \le i \le p$.
Then, a completely positive map $\sum_{i=1}^p b^*_i \cdot T_i \cdot b_i$ may be defined on
$M_n(\bC)$ by

\begin{equation}
(\sum_{i=1}^p b^*_i \cdot T_i \cdot b_i)(a) \equiv \sum_{i=1}^p b^*_i T_i(a) b_i
\end{equation}

for all $a \in M_n(\bC)$.

\begin{definition}
 A subset $\cV \subset \C \cP$ is called matricially convex {\rm(}see \cite{Arverson}, 
\cite{Paulsen}{\rm )} if for $T_i \in \cV$, $1 \le i \le p$ and for 
all $b_i \in M_n(\bC) $ such that $\sum_{i=1}^p b^*_ib_i = \jed$
it follows that
$$\sum_{i=1}^p b^*_i \cdot T_i \cdot b_i \in \cV.$$
\end{definition}
 One has

\begin{theorem}[Smith, Ward \cite{SW}]
There is a one-to-one correspondence between matricially convex faces in $\C \cP$ and faces in the state 
space $\cS$.
\end{theorem}

Now it is clear that this result combined with Theorem \ref{AlfsenShultz} says that {\bf there
are ``more'' faces in $\C\cP$ than projectors in $M_n(\bC)$}. Hence, it is natural to restrict the 
class of faces which we are interested in. Following this idea we
turn to a characterization of all maximal faces of $\cP$ and $\C\cP$.
We begin with

\begin{proposition}(Kye \cite{Kyecan})
\label{maximalface}
Every maximal face of $\cP$ is of the form
\begin{equation}
\label{4.2}
F_{max}(p_{\xi}, \eta) = \{ T \in \cP; T(p_{\xi}) \eta =0 \}.
\end{equation}
where $p_{\xi}$ is a one dimensional projection on $\xi$ and $\eta$ is another nonzero vector.
Moreover, if $F_1$ and $F_2$ are two maximal faces of $\cP$ then they 
are affine isomorphic to each other.
\end{proposition}

Consequently, any maximal face of $\cP$ corresponds to a pair of one dimensional subspaces in
$\bC^n$.
The maximal faces in $\C \cP$ are characterized by

\begin{proposition}(Kye \cite{nowyKye})
\label{maximalfaces2}
Every maximal face of $\C \cP$ is of the form
\begin{equation}
\label{4.3}
F_{max}(V) = \{ T_{\cW} \in \C\cP; \cW \subset V^{\perp} \}.
\end{equation}
where $V \in M_n(\bC)$ and $\perp$ is understood in the sense of the inner product
$<V,W> = Tr(W^*V)$.  $T_{\cW}$ and $\cW$ are defined via relation (\ref{3.1}) with
$\tau_i$ $^*$-homomorphisms.
\end{proposition}

Consequently, in the finite dimensional case, {\bf there is a complete characterization
of maximal faces in $\cP$ and $\C\cP$}. Moreover, in these cases every face of $\cD$ is 
the convex hull of a face of $\C\cP$ and a face of completely copositive maps \footnote{Completely
copositive map is the composition of transposition with a CP map.} $co-\C\cP$ 
(cf \cite{Kyedecom}). 
Hence, also one has as a corollary

\begin{corollary}
$\cD$ is a convex hull of $\{ F_{max}(V_1), \tau \circ F_{\max}(V_2); V_1, V_2 \in M_n(\bC)\}$
where $\tau$ stands for the transposition.
\end{corollary}

The relation between maximal faces (\ref{4.2}) of $\cP$ and maximal faces (\ref{4.3}) of $\C\cP$ is given by

\begin{proposition}(Kye \cite{nowyKye})
\label{maximalfaces3}
Let $V = |\xi><\eta|$ with unit vectors $\xi, \eta \in \bC^n$.
Then one has the identity
\begin{equation}
F_{max}(V) = F_{max}(p_{\xi}, \eta) \cap \partial \C\cP.
\end{equation}
where $\partial \C\cP$ stands for the boundary of $\C\cP$. Moreover, for such $V$
\begin{equation}
F_{max}(V) \subseteq F_{max}(p_{\xi}, \eta).
\end{equation}
\end{proposition}

Again, there is a more specialized result, see \cite{K1}. Namely, denote by $\cP_k$
the convex cone of all $k$-positive maps from $M_n(\bC)$ into $M_n(\bC)$. Kye has shown that
every maximal face of $\cP_k$ corresponds to an $n \times n$ matrix whose rank
is less or equal to $k$. Hence, the number of maximal faces of $\cP_k$  grows
with $k$. However, the number of maximal faces of $\cP_k$ which are contained in the 
boundary of $\cP$ is constant and determined by matrices of rank one (see Corollary 3.2 in \cite{K1}).
 We end this section with another Kye's result:

\begin{proposition}(Kye \cite{Kyecan})
For a positive linear map $T \in \cP$, the following are equivalent:
\begin{itemize}
\item{} $T$ is an interior point of $\cP$.
\item{} $T(p_{\xi})$ is nonsingular for each one-dimensional projection $p_{\xi} \in \cB(\bC^n)
\equiv M_n(\bC)$.
\end{itemize}
\end{proposition}

Consequently, interior points of $\cP$ are ``far'' from $\Fs$.

\section{POSITIVE MAPS AND LOCALLY DECOMPOSABLE MAPS}
In this Section we outline briefly the general construction of a linear positive map
$T : \cB(\cH) \to \cB(\cK)$ with an emphasis on the local decomposability and extreme positive maps.
Here, again, $\cH$ and $\cK$ are finite dimensional Hilbert spaces of dimension greater than 1.

For any $x \in \cH$ we define the linear operator $V_x : \cK \to \cH \otimes \cK$
by $V_xz = x \otimes z$ for $z \in \cK$. By $e_{x,y}$ ,where $x,y \in \cH$, we denote the one dimensional 
operator on $\cH$ defined by $e_{x,y}u = (y,u)x$ for $u \in \cH$, i.e. $e_{x,y} \equiv |x><y|$.
For simplicity reasons, if $\{v_i\}_1^n$ is a basis in $\cH$, we will write $V_i$ and
$e_{i,j}$ instead of $V_{v_i}$ and $e_{v_i,v_j}$ for any $i,j = 1,2,...,n$ when no confusion can arise.

Let $H \in \cB(\cH \otimes \cK)$. Define $T_H: \cB(\cH) \to \cB(\cK)$ as follows
\begin{equation}
T_H(e_{x,y}) = V^*_x H V_y,
\end{equation}
where $x,y \in \cH$. It was Choi, \cite{Choi}, who firstly discovered correspondences among 
various types of 
$H \in \cB(\cH \otimes \cK)$ and classes of linear positive maps $T_H$ 
(see also \cite{Jam}). We will need the following result
(cf. \cite{Jam}, \cite{MM})
\begin{theorem}
\label{5.1}
If $H = H^*$ and $(x \otimes y, H x \otimes y) \ge 0$ for any $x \in \cH$ and $y \in \cK$  then $T_H$ is a positive map. 
Moreover, for any positive map $T: \cB(\cH) \to \cB(\cK)$ there exists uniquely determined selfadjoint 
operator $H \in \cB(\cH \otimes \cK)$ with the property $(x \otimes y, H x \otimes y) \ge 0$ 
for any $x \in \cH$ and $y  \in \cK$, such that
$T = T_H$. 
\end{theorem}

It should be mentioned that Choi \cite{Choi} proved the following remarkable result 
concerning complete positive maps:
{\it $T_H$ is a completely positive map if and only if $H$ is a positive operator}; so not 
only ``block-positive`` as in Theorem(\ref{5.1}).
 
\smallskip

The important point to note here is that there is an explicit relation between $H$ and $T$.
Namely, (cf. \cite{Jam}, \cite{MM}) suppose $T: \cB(\cH) \to \cB(\cK)$ is any positive map and define
\begin{equation}
\label{5.2}
H = (\id \otimes T) \bigl( \sum_{kl} |\xi_k><\xi_l| \otimes |\xi_k><\xi_l| \bigr),
\end{equation}

 for a basis $\{\xi_j\}$ in $\cH$. For any $y,w \in \cK$ we have
\begin{equation}
(y, T_H(e_{ij})w) = (y, V^*_{\xi_i} H V_{\xi_j}w) = \sum_{kl} (\xi_i , |\xi_k><\xi_l| \xi_j)
(y, T(|\xi_k><\xi_l|w) = (y, T(e_{ij})w),
\end{equation}
where $e_{ij} \equiv |\xi_i><\xi_j|$.

In the sequel, we will need another very important property of positive maps. 
This property, called local decomposability,
is defined as follows (cf \cite{St}):

\begin{definition}
\label{locallydecom}
A linear map $\tau : \cB(\cH) \to \cB(\cH)$
is locally decomposable
if for $0 \ne x \in \cH$, there exists a Hilbert space $\cK_x$, 
a bounded operator $W_x:\cK_x \to \cH$
and a \Cs-homomorphism (equivalently, Jordan homomorphism) $\pi_x$ of $\cB(\cH)$ to $\cB(\cK_x)$ such that  
\begin{equation}
\label{stormer}
W_x \pi_x(a) W^*_x x = \tau (a)x,
\end{equation}
for all $a \in \cB(\cH)$. 
\end{definition}

It was St{\o}rmer who proved

\begin{theorem}{\rm (}St{\o}rmer \cite{St}{\rm )}
\label{locallydecomtw}
Every bounded positive linear map of a \Cs-algebra $\cA$ into the bounded
operators on a Hilbert space $\cH$ is locally decomposable. Moreover, if $a$ in
\ref{stormer} is selfadjoint, then $\pi$ (in \ref{stormer}) can be taken to be 
$^*$-morphism.
\end{theorem}

Hence, every positive linear map is locally decomposable, but in 2D-case (two dimensional)
the notions of decomposability and local decomposability are the same as exactly for this case
every positive map is decomposable one, see \cite{Wor}, \cite{St}. Going to higher dimensions, so for nD-case 
(n dimensional) with $n>2$,
there are non-decomposable maps which are only locally decomposable.
This explains our remark given in Introduction that properties of positive maps for 2-level 
systems and 3-level systems are dramatically different. 
We claim that to understand this difference we should use the facial geometry of 
the underlying convex structures (presented above). 
This will be done in the next Section.

However, we want to close this Section with another St{\o}rmer's result. He obtained in 2D-case 
(and only for this case) the classification of all extreme points in $\cP$.

\begin{theorem} {\rm (} St{\o}rmer \cite{St} {\rm )}
\label{extremalne}
Let $T : M_2(\bC) \to M_2(\bC)$ be a positive map. Then $T$ is extreme if and 
only if $T$ is unitarily equivalent to a map of the form
\begin{equation}
\label{5.5}
\left( \begin{array}{cccc} 
a &     b \\    
c &  d  \end{array} \right)
\mapsto
\left( \begin{array}{cccc} 
a                           &                                   \alpha b + \beta c\\    
\overline{\alpha}c + \overline{\beta}b & \gamma a + \epsilon b +\overline{\epsilon} c + \delta d  \end{array} \right)
 \end{equation}

where $|\epsilon|^2 = 2 \gamma (\delta - |\alpha|^2 - |\beta|^2)$ in the case  when
$\gamma \ne 0$, and $|\alpha|$ or $|\beta|$ equals 1 when $\gamma = 0$.
In the former case, $|\alpha| + |\beta| = \delta^{1/2}$.
\end{theorem}

\section{LOW DIMENSIONAL CASES:  $M_2(\bC)\quad$ and $ \quad M_3(\bC)$}

To understand the phenomenon of non-decomposable maps we should 
firstly recognize the meaning of 
locally decomposable maps, see Definition \ref{locallydecom} and Theorem \ref{locallydecomtw}.
To this end we will compare the facial structure
of 2D-case with that for 3D-case. Let us start with 2D-case. The maximal faces of $\cP$ are 
characterized by Proposition \ref{maximalface}. We wish to combine the general form of a maximal 
face of $\cP$ with the local decomposability. Assume that a unital positive map $T$ is in
a fixed arbitrary maximal face, i.e. $T \in \Fs$. Define a functional
$\phi_T: M_2(\bC) \to \bC$ such that 
\begin{equation}
\phi_T(\cdot) = (\eta, T(\cdot) \eta).
\end{equation}
  Following the GNS recipe one has
\begin{equation}
\cH_{\phi_T} = M_2(\bC)/\cL_{\phi_T}
\end{equation}
where $\cL_{\phi_T} = \{ a \in M_2(\bC): \phi_T(a^*a) =0 \} = M_2(\bC) p$  
for an orthogonal projector $p \in M_2(\bC)$.
The definition of $\Fs$ implies that $p = p_{\xi}$. 
Hence $\cH_{\phi_T} = \cH_{\phi_{T^{\prime}}}$ provided that $T, T^{\prime} \in \Fs$.
Furthermore, the $C^*$-homomorphism $\pi$ (cf. Definition \ref{locallydecom})
is the same map for all positive maps in the fixed face.
More precisely, one can define $ \cR_{\phi_T} =\{a\in M_2(\bC):\, \phi_T(aa^*)=0\}$.
$\cR_{\phi_T}$ is a right ideal. By $\cH_{\phi_T}^0$ we denote the quotient space $M_2(\bC)/\cR_{\phi_T}$.
For any $a\in M_2(\bC)$ we write $[a]_{\rm l}$ and $[a]_{\rm r}$ the equivalence classes (see A20)
 of $a$ in  
$\cH_{\phi_T}$ and $\cH_{\phi_T}^0$
respectively. For simplicity we will write $[a]$ instead of $[a]_{\rm l}\oplus[a]_{\rm r}$ 
for $a\in M_2(\bC)$.
Next, let $\cK_\eta= \cH_{\phi_T}\oplus \cH_{\phi_T}^0$. $W_\eta$ and $\pi_\eta$ are
given by
\begin{equation}\label{rho}
\pi_{\eta}(a)\left([b_1]_{\rm l}\oplus[b_2]_{\rm r}\right)=[ab_1]_{\rm l}\oplus
[b_2a]_{\rm r},\;\;\;a,b_1,b_2\in M_2(\bC);
\end{equation}
\begin{equation}\label{V}
W_\eta \pi_{\eta}(a)[\jed] = T(a)\eta.
\end{equation}

Consequently,  we are able to write all ingredients of local decomposability in explicit way. 
However, to obtain decomposability 
within the St{\o}rmer construction one should add the additional condition (see \cite{MM1}).
To present this result we 
need some notations. If $\xi$ and $\eta$ are arbitrary unit vectors in $\bC^2$ then
let $\xi_1,\xi_2$ be an orthonormal basis in $\bC^2$ such that $\xi_1=\xi$, $\xi_2 = \xi^{\perp}$ 
and similarly 
$\eta_1,\eta_2$ be a basis
such that $\eta_1=\eta$. Again, by $e_{ij}$ we denote the operator $|\xi_i> <\xi_j|$ for $i,j=1,2$.
\begin{proposition}
\label{stwierdzenie1}
Suppose  a  unital positive map $T \in \Fs$. Let $\cK_\eta$,
$W_\eta$ and $\pi_\eta$ be as in (\ref{stormer}) (and described by \ref{rho} - \ref{V}). Then 
the condition for decomposability 
\begin{equation}\label{main}
T(a)=W_\eta\pi_\eta(a)W_\eta^*,\;\;\; a\in M_2(\bC)
\end{equation}
is satisfied if and only if
\begin{equation}\label{tr}
\Tr\{T(e_{12})\}=\Tr\{T(e_{21})\}=0,\;\;\;
\Tr\{T(e_{22})\}=1,
\end{equation}
\begin{equation}\label{alfabeta}
\Tr\{T(e_{11})\}=2\left(|<\eta_2,T(e_{12})\eta_1>|^2+|<\eta_2,
T(e_{21})\eta_1>|^2\right).
\end{equation}
\end{proposition}

This result clearly shows that even in the simple 2D case, local decomposability does not lead directly
to decomposability (we recall that in 2D case each positive map is decomposable).
However, for the considered case one can go one step further (see \cite{MM1}).
Namely, easy calculations lead to
the explicit form $H_T = \sum_{i,j} e_{ij} \otimes T(e_{ij})$ (cf. (\ref{5.2})) in the 
basis $\{ \xi_i \otimes \eta_k \}$. One has

\begin{equation}
\label{odwzorowanie}
H_T =
\left( \begin{array}{cccc} 
0  &     0        &      0      &      y\\    
0  & \lambda   & \overline{z}   &    t   \\
0  &z          &       1        &   0    \\
\overline{y} & \overline{t} &   0   &  1 - \lambda  \end{array} \right)
 \end{equation}

where $\lambda \in [0,1]$ and for any $x,v \in \bC^2$

\begin{eqnarray}
\label{6.9}
\lefteqn{\lambda |(\xi_1,x)|^2 |(v,\eta_2)|^2 + |(\xi_2,x)|^2 |(v, \eta_1)|^2}
                                                                             \nonumber\\
\lefteqn{ + (1 - \lambda)|(\xi_2,x)|^2 |(v, \eta_2)|^2 +2 Re \{(x,\xi_1)(\xi_2,x)|(v, \eta_2)|^2 t \} }
                                                                                    \nonumber\\
& & \ge - 2 Re\{ (x,\xi_1)(\xi_2,x)[ y (v, \eta_1)(\eta_2, v) + \overline{z}(v,\eta_2)(\eta_1,v)] \}.
\end{eqnarray}

Moreover, these calculations give the following explicit form of a map in the maximal face:
\begin{eqnarray}
\label{odwzorzesciany}
T(|\xi_1><\xi_1|)  & = & \lambda |\eta_2><\eta_2|,  \\
T(|\xi_1><\xi_2|)  & = & y |\eta_1><\eta_2| + \overline{z} |\eta_2><\eta_1| + t |\eta_2><\eta_2|, \\
\label{odwzorzesciany1}
T(|\xi_2><\xi_2|)  & = & |\eta_1><\eta_1| + (1 - \lambda) |\eta_2><\eta_2|.
\end{eqnarray}

where, we recall,  $\xi_1 \equiv \xi$, $\xi_2 \equiv \xi^{\perp}$, anologously for $\eta$'s.
Numbers $\lambda$, $z$, $y$, and $t$ satisfy a condition of the type (\ref{6.9}).

The important point to note here is the rather striking similarity between (\ref{odwzorowanie})
and the St{\o}rmer result (\ref{5.5}). Namely, the Choi's matrix for extreme positive map
has the form

\begin{equation}
\label{6.10}
\left( \begin{array}{cccc} 
1  &     0        &      0      &      \alpha \\    
0  & \gamma   & \overline{\beta}   &    \epsilon   \\
0  & \beta          &       0        &   0    \\
\overline{\alpha} & \overline{\epsilon} &   0   &  \delta  \end{array} \right)
 \end{equation}
obviously, with the same conditions for $\alpha, \beta, \gamma, \delta, \epsilon$ 
as these stated in Theorem \ref{extremalne}. 

Secondly, we note that LHS(\ref{6.9}) does not depend on phases of the complex numbers $(v,\eta_k)$, $k=1,2$
while RHS(\ref{6.9}) does.  
In particular, there are many vectors $v \in \bC^2$ 
with the property that the coefficient of $z$ ($y$ respectively) in RHS(\ref{6.9}) is equal to $0$.
This suggests the possibility of splitting the family of matrices 
(\ref{odwzorowanie}) into two classes

\begin{equation}
\label{odwzorowanie1} 
\left( \begin{array}{cccc} 
0  &     0        &      0      &      y\\    
0  & \lambda^{\prime}   &  0   &    t^{\prime}   \\
0  &     0      &       q^{\prime}        &   0    \\
\overline{y} & \overline{t^{\prime}} &   0   &  \frac{1}{2} - \lambda^{\prime}  \end{array} \right)
 \end{equation}
with $\lambda^{\prime} \in [0,1]$ and for any $x,v \in \bC^2$

\begin{eqnarray}
\label{6.9a}
\lefteqn{\lambda^{\prime} |(\xi_1,x)|^2 |(v,\eta_2)|^2 + q^{\prime}|(\xi_2,x)|^2 |(v, \eta_1)|^2
+(\frac{1}{2} - \lambda^{\prime})|(\xi_2,x)|^2 |(v, \eta_2)|^2 }
                                         \nonumber\\
& & +2 Re \{(x,\xi_1)(\xi_2,x)|(v, \eta_2)|^2 t^{\prime} \} 
\ge - 2 Re\{ (x,\xi_1)(\xi_2,x) y (v, \eta_1)(\eta_2, v) \}
\end{eqnarray}

and
\begin{equation}
\label{odwzorowanie2}
\left( \begin{array}{cccc} 
0  &     0        &      0      &      0\\    
0  & \lambda''   & \overline{z}   &    t''   \\
0  & z          &       q''        &   0    \\
0    & \overline{t''} &   0   &  \frac{1}{2} - \lambda''  \end{array} \right)
 \end{equation}
where $\lambda'' \in [0,1]$, for any $x,v \in \bC^2$
\begin{eqnarray}
\label{6.9b}
\lefteqn{\lambda'' |(\xi_1,x)|^2 |(v,\eta_2)|^2 + q''|(\xi_2,x)|^2 |(v, \eta_1)|^2
+(\frac{1}{2} - \lambda'')|(\xi_2,x)|^2 |(v, \eta_2)|^2 }
                                            \nonumber\\
& & +2 Re \{(x,\xi_1)(\xi_2,x)|(v, \eta_2)|^2 t'' \}
 \ge - 2 Re\{ (x,\xi_1)(\xi_2,x)\overline{z}(v,\eta_2)(\eta_1,v) \},
\end{eqnarray}
and $\lambda^{\prime} + \lambda'' = \lambda$, $t^{\prime} + t'' = t$,
$q^{\prime} + q'' = 1$.

The maps determined by matrices of the form (\ref{odwzorowanie2}) have a very interesting property.
To describe this feature of the corresponding positive maps we recall
Choi's result saying (see Section 5) that a map determined by a positive matrix is completely 
positive, i.e. $T$ is a completely positive map if and only if the $2 \times 2$ operator matrix

\begin{equation}
\label{AndoChoi}
\left( \begin{array}{cccc} 
T(e_{11})  &    T(e_{12})\\    
T(e_{21})  &  T(e_{22}) \end{array} \right)
 \end{equation}

is positive. We recall $e_{ij} \equiv |\xi_i><\xi_j|$. On the other hand, it is 
well-known \cite{Ando}, \cite{Choi3} (see also \cite{HJ} where the matrix version of this 
inequality is described) that any matrix of 
the form (\ref{AndoChoi}) is positive if and only if $T(e_{11})$, $T(e_{22})$ are positive
and $T(e_{11}) \ge T(e_{12}) T(e_{22})^{-1} T(e_{12})^*$. Here if $T(e_{22})$ is not invertible
$T(e_{22})^{-1}$ is understood to be its generalized inverse. The latter, Ando-Choi, inequality leads to
the following condition on $\lambda''$, $z''$ and $t''$:
\begin{equation}
\label{6.19AC}
\lambda'' \ge (q'')^{-1}|z|^2 + (\frac{1}{2} - \lambda'')^{-1} |t''|^2.
\end{equation}

On the other hand, Corollary 8.4 in \cite{St} implies that the map $T_H$ (i.e. the map determined by 
the matrix $H$ of the form \ref{odwzorowanie2}) is positive if and only if
\begin{equation}
| \overline{z} (x, \eta_2)(\eta_1,x) +t'' |(\eta_2,x)|^2|^2 \le 
\lambda'' |(\eta_2,x)|^2 ( q''|(\eta_1,x)|^2  + (\frac{1}{2} - \lambda'') |(\eta_2,x)|^2),
\end{equation}
for any $x \in \bC^2$.
In particular
\begin{equation}
\bigl( |(\eta_1,x)| |z| +|(\eta_2,x)| |t''| \bigr)^2 \le \lambda''(q''|(\eta_1,x)|^2 
+ (\frac{1}{2} - \lambda'')|(\eta_2,x)|^2).
\end{equation}

Without loss of generality we can assume $|(\eta_1,x)| \ne 0$.
Let us define $\sigma = \frac{|(\eta_2,x)|}{|(\eta_1,x)|}.$
Then
\begin{equation}
(|z| + \sigma |t''|)^2 \le  \lambda''(q'' +(\frac{1}{2} - \lambda'') \sigma^2)
\end{equation}
so
\begin{equation}
\label{6.18}
(|t''|^2 - \lambda''(\frac{1}{2} - \lambda'')) \sigma^2 +2 |t''| |z| \sigma + |z|^2 - \lambda'' q'' \le 0.
\end{equation}

The only admissible case is when the discriminant of the quadratic equation (\ref{6.18})
is negative, i.e. $\Delta \le 0$.
However, this implies
\begin{equation}
q''|t''|^2 +(\frac{1}{2} - \lambda'')|z|^2 \le q''\lambda''(\frac{1}{2} - \lambda'').
\end{equation}
But this means (cf \ref{6.19AC}) that for the studied class of maps, positivity implies complete positivity.

\smallskip

Now, let us turn to maps determined by matrices of the form (\ref{odwzorowanie1}).
The first easy observation says that an application of partial transposition to matrices of the form 
(\ref{odwzorowanie1}) leads to matrices of the form (\ref{odwzorowanie2}).
But then combining the argument given in the preceding paragraph with the relation between the matrix $H$
and the positive map $T_H$ given in Section 5 one can conclude that matrices of the form 
(\ref{odwzorowanie1}) correspond to co-completly positive maps. Therefore, the considered splitting
of matrix (\ref{odwzorowanie}) corresponds to decomposition of a positive map into the sum of 
completely positive and completely co-positive maps provided that conditions \ref{6.9a} and 
\ref{6.9b} are satisfied. 

\smallskip 

Consequently, any unital positive map $T$ in the face $\Fs$ (cf Section 5) is decomposable one 
and {\it the decomposition can be written explicitly}. 
Clearly, this extends to a map $S \in \cP$ since such $S$ is a convex combination of maps having the
form \ref{odwzorzesciany} - \ref{odwzorzesciany1}.
The important point to note here is the form of maps which constitute the discussed decomposition:
{\bf both are not normalized,} i.e. the summands do not preserve the identity. However, the summands are 
in the same face. Consequently, we got an indication that for explicit splitting of decomposable 
map the face structure is appearing as the natural one.

\smallskip

The presented decomposition of positive maps for 2D-case is not conclusive as there is still one 
unanswered question whether condition \ref{6.9} implies \ref{6.9a} and \ref{6.9b} (the converse 
implication is easy). In other words, we wish to decompose any matrix (\ref{odwzorowanie}) satisfying 
the general conditions (\ref{6.9}). This can be done but, as we just learnt, there is a price to pay. Namely,
the normalization is lost, i.e. in general, the summands in the decomposition
do not preserve identity.

More precisely, we will show that (\ref{6.9}) implies (\ref{6.9a}) and (\ref{6.9b}). To this end,
multiplying (\ref{6.9}) by $|y|\lambda^{-\frac{1}{2}}$ 
and assuming that $|y| + |z| = \lambda^{\frac{1}{2}}$ (this is the property characterizing 
an extremal positive 
map, cf Theorem \ref{extremalne})  one can show that the matrix

\begin{equation}
\label{odwzorowanie1a}
H_{T_1} =
\left( \begin{array}{cccc} 
0  &     0        &      0      &      y\\    
0  & \lambda_1   &  0   &    t_1   \\
0  &  0          &       a_1        &   c   \\
\overline{y} & \overline{t_1} &   \overline{c}   &  b_1  \end{array} \right)
 \end{equation}

corresponds to the positive map $T_1$, as the following condition holds

\begin{eqnarray}
\label{6.10a}
\lefteqn{\lambda_1 |(\xi_1,x)|^2 |(v,\eta_2)|^2 + a_1|(\xi_2,x)|^2 |(v, \eta_1)|^2}
                                                                             \nonumber\\
\lefteqn{ + b_1|(\xi_2,x)|^2 |(v, \eta_2)|^2 +2 Re \{(x,\xi_1)(\xi_2,x)|(v, \eta_2)|^2 t_1 \} }
                                                                                    \nonumber\\
& & \ge - 2 Re\{ (x,\xi_1)(\xi_2,x) y (v, \eta_1)(\eta_2, v) + c |(x, \xi_2)|^2 (v,\eta_1)(\eta_2,v) \}.
\end{eqnarray}

where  we put $a_1 = |y| \lambda^{-{1 \over 2}}$, $\lambda_1 = |y| \lambda^{1\over 2}$, 
$t_1 = {t \over 2}$,
$b_1 = |z| \lambda^{-{1 \over 2}} (1 - \lambda)$, and finally $c \in \bC$.

Repeating this argument, i.e.  multiplying (\ref{6.9}) by $|z|\lambda^{-{1\over 2}}$ 
and again assuming that $|y| + |z| = \lambda^{1 \over 2}$  one can show that the matrix

\begin{equation}
\label{odwzorowanie2a}
H_{T_2} =
\left( \begin{array}{cccc} 
0  &     0        &      0      &      0\\    
0  & \lambda_2   & \overline{z}   &    t_2   \\
0  &z          &       a_2        &   -c    \\
0 & \overline{t_2} &  - \overline{c}   &  b_2 \end{array} \right)
 \end{equation}

corresponds to the positive map $T_2$, as the following condition holds

\begin{eqnarray}
\label{6.10b}
\lefteqn{\lambda_2 |(\xi_1,x)|^2 |(v,\eta_2)|^2 + a_2 |(\xi_2,x)|^2 |(v, \eta_1)|^2}
                                                                             \nonumber\\
\lefteqn{ + b_2 |(\xi_2,x)|^2 |(v, \eta_2)|^2 +2 Re \{(x,\xi_1)(\xi_2,x)|(v, \eta_2)|^2 t_2 \} }
                                                                                    \nonumber\\
& & \ge - 2 Re\{ (x,\xi_1)(\xi_2,x) \overline{z}(v,\eta_2)(\eta_1,v) - c |(x, \xi_2)|^2 (v, \eta_1)
(\eta_2, v)\}
\end{eqnarray}

where  we put $a_2 = |z| \lambda^{-{1 \over 2}}$, $\lambda_2 = |z| \lambda^{1\over 2}$, 
$t_2 = {t \over 2}$,
$b_2 = |y| \lambda^{-{1 \over 2}} (1 - \lambda)$, and finally $c \in \bC$.  

Clearly, $\lambda_1 + \lambda_2 = \lambda$, $a_1 + a_2 = 1$, $t_1 + t_2 = t$ , and
$b_1 + b_2 = 1 - \lambda$. Consequently, \ref{odwzorowanie1a} and \ref{odwzorowanie2a}
gives the desired splitting of \ref{odwzorowanie}. Furthermore, if $c$ satisfies 

\begin{equation}
\overline{y_1} \lambda^{-{1 \over 4}} t_1 = y_1 \lambda^{ 1\over 4} \overline{c},
\end{equation}
where $y_1^2 = y$ then 
the matrix \ref{odwzorowanie2a}
is positive, thus $T_2$ is CP map. Similarly, for the special choice of $c$ , the matrix
\ref{odwzorowanie1a} corresponds to co-CP map.
As a result, whenever $\lambda, y, z$ are not equal to $0$
we obtained {\bf the unique decomposition} of an extremal positive map into the sum of CP
and co-CP maps but {\bf both, in general, are not normalized}. This is indicated by the fact that
both matrices contain, in general,  $c \ne 0$, and the sum of diagonal elements does not
need to be 2.

\smallskip

Now, before turning to $3D$-case, let us consider unital positive maps from 
$M_2(\bC) \to M_3(\bC)$. Again, our starting point is the explicit form of maximal faces 
$F_{max}(p_{\xi}, \eta)$ in $\cP$ (cf Section 4).  Let us take a unital positive map
$ T \in F_{max}(p_{\xi}, \eta)$ and pick up two bases $\{ \xi_k \}_{k=1}^2$ and $\{ \eta_l \}_{l=1}^3$ 
in $\bC^2$ and $\bC^3$ such that $\xi_1 \equiv \xi$ and $\eta_1 \equiv \eta$ respectively.
We observe
\begin{equation}
\label{6.23}
\eta_1 = T(\jed) \eta_1 = \sum_{k=1}^2 T(p_{\xi_k}) \eta_1 = T(p_{\xi_2})\eta_1 .
\end{equation}

We can conclude from (\ref{6.23}) as well as from the given description of maximal face that
the explicit form of Choi's matrix $H_T$ (for the normalized map T) is:

\begin{equation}
\label{odwzorowanie3d}
H_{T} =
\left( \begin{array}{ccccccc}
0  &     0  & 0 & 0     &      v_{12}      &     v_{13}\\    
0  &    a &  c   &    v_{21} &   v_{22}   &   v_{23}\\
0  &  \overline{c}  &  b  &  v_{31}  & v_{32}    &  v_{33}\\
0 & \overline{v_{21}} & \overline{v_{31}} &  1 & 0 & 0\\ 
\overline{v_{12}} & \overline{v_{22}} & \overline{v_{32}} & 0 & 1 - a & -c\\
\overline{v_{13}} & \overline{v_{23}} & \overline{v_{33}} & 0 & - \overline{c} & 1 - b\end{array} \right)
\equiv 
\left( \begin{array}{cccc} 
A_{11}  &    A_{12}\\    
A_{21}  &  A_{22} \end{array} \right)
 \end{equation}

where $a,b$ are non-negative numbers, $c, v_{ij}$ are in $\bC$, $v_{11} = 0$ due to the block-positivity,
 $|c|^2 \le ab$ and $v_{ij}$ 
satisfy the condition of the 
type \ref{6.9} (now a little bit complicated). Finally, the last equality says that we partitioned 
the matrix $H_T$, i.e. 
$H_T$ is written as $2 \times 2$ square matrix with entries $A_{ij} \in M_3(\bC)$. 
We note that $A_{21} = A_{12}^*$.
We recall that the decomposability of any unital positive map holds for this case(see \cite{Wor}).
Futhermore, in terms of Choi's matrix, it means that there
are two block-matrices splitting \ref{odwzorowanie3d}. The first matrix has the form

\begin{equation}
\label{2na3,a}
\left( \begin{array}{cccc} 
A^I_{11}  &    A^I_{12}\\    
A^I_{21}  &  A^I_{22} \end{array} \right)
\end{equation}
where $A_{11}^I$ and $A_{22}^I$ are positive semidefinite matrices with $A_{12}^I$ satisfying

\begin{equation} 
|(x,A_{12}^I y)| \le ||(A_{11}^I)^{\frac{1}{2} } x|| \cdot ||(A_{22}^I)^{\frac{1}{2} } y||,
\end{equation}
for any $x,y \in \bC^3$. The second block-matrix is of the form

\begin{equation}
\label{2na3,b}
\left( \begin{array}{cccc} 
A^{II}_{11}  &    A^{II}_{12}\\    
A^{II}_{21}  &  A^{II}_{22} \end{array} \right)
\end{equation}
where $A_{11}^{II}$ and $A_{22}^{II}$ are positive semidefinite matrices. 
Furthermore, $A_{12}^{II}$ satisfies

\begin{equation} 
|(x,(A_{12}^{II})^* y)| \le ||(A_{11}^{II})^{\frac{1}{2} } x|| \cdot ||(A_{22}^{II})^{\frac{1}{2} } y||,
\end{equation}
for any $x,y \in \bC^3$, and $A_{ij} = A^I_{ij} + A^{II}_{ij}$.

Now, we are in position to consider 3D-case. Let us consider a unital positive map
$T \in F_{max}(p_{\xi}, \eta)$ and 
 pick up two bases $\{ \xi_k \}_{k=1}^3$ and $\{ \eta_l \}_{l=1}^3$ 
in $\bC^3$ such that $\xi_1 \equiv \xi$ and $\eta_1 \equiv \eta$ respectively.
We observe
\begin{equation}
\label{6.23b}
\eta_1 = T(\jed) \eta_1 = \sum_{k=1}^3 T(p_{\xi_k}) \eta_1 = T(p_{\xi_2})\eta_1 
+ T(p_{\xi_3}) \eta_1.
\end{equation}
We can only conclude from (\ref{6.23b}) 
that $T(p_{\xi_k})$, $k=2,3$ are positive operators
in $\cB(\bC^3)$ such that their sum $T(p_{\xi_2}) + T(p_{\xi_3})$ has $\eta_1$ as its eigenvector.
The Choi's matrix $H_T$ (for the considered map T) is given by:

\begin{equation}
\label{odwzorowanie3D}
H_{T} =
\left( \begin{array}{cccc} 
A_{11}  &   A_{12}  & | &   A_{13}\\    
A_{21}  &   A_{22}  & | &   A_{23}\\
--      &   --      & | &   --   \\
A_{31}  &   A_{32}   & | &    A_{33}\end{array} \right)
 \end{equation}

where $A_{ij}$, $i,j = 1,2,3$, are $3 \times 3$ matrices such that

\begin{equation}
\label{odwzorowanie3D1}
\left( \begin{array}{cccc} 
(x,A_{11}x) & (x, A_{12}x) & (x,A_{13}x)\\    
(x,A_{21}x) & (x,A_{22}x)  & (x,A_{23}x)\\
(x,A_{31}x) & (x,A_{32}x)  & (x, A_{33}x) \end{array} \right)
 \end{equation}

is positive semidefinite matrix for any $x \in \bC^3$ 
(in particular, $A_{kk}$ are positive semidefinite matrices).
The formulae \ref{odwzorowanie3D} and \ref{odwzorowanie3d} 
suggest that the method only based on matricial analysis
of the Choi operator becomes too complicated to be effectively used for undestanding the nature
of non-decomposable maps; there are too many variables. Hence, we will exploit the geometrical structure
described in Sections 2 and 3. To this end we begin with remark that
the indicated partitioning of the matrix (\ref{odwzorowanie3D}) corresponds to the separation of
the matrix $\{ A_{ij} \}_{i,j=1}^2$ which can be attributed to 
family of all maps: $M_2(\bC) \to M_3(\bC)$. As it was mentioned, unital positive maps:
$M_2(\bC) \to M_3(\bC)$ are decomposable with the transposition associated to the basis
$\xi_i \otimes \eta_k$, where $i=1,2$ and $k=1,2,3$. If one could expect decomposability for 
3D-case (we know that there are counterexamples, see \cite{Choi}, \cite{Wor}) the corresponding
transposition would be associated with the basis 
$\xi_i \otimes \eta_k$, where $i=1,2,3$ and $k=1,2,3$. In general these 
transpositions do not need to co-operate well. 
Hence, one guesses that decomposable maps are,
 somehow, more regular than plain positive maps. Let us examine this question in detail.

A decomposable map $\alpha: \cB(\cH) \to \cB(\cH)$ can be written as 
the composition of a Jordan morphism $\tau :
\cB(\cH) \to \cB(\cK)$ 
with the map $U_W$, $U_W(b) \equiv W^*bW$, $b \in \cB(\cK)$ where $W:\cH \to \cK$,
i.e. $\alpha = U_W \circ \tau$ (see also Section 3)\footnote{At first sight this form of decomposable maps
can be taken as the particular case of definition given in Section 3. However, this is not true as 
the Hilbert space $\cK$ can be taken large enough to take  into account all summands given in 
right hand side 
of formula \ref{3.1}.}.  
We know that Jordan morphisms are regular in the sense that they respect certain properties
of the facial
structure, see Theorem \ref{ASJor} and Corollary \ref{wniosek1}. Thus, 
to examine the regularity of decomposable maps we should carefully study 
the composition of $U_W$-maps with Jordan morphisms.
Denote $\tau(\jed)$ by $Q \in \cB(\cK)$. As Jordan morphisms send (orthogonal) projectors 
into (orthogonal) projectors, $Q$ is a projector. We put $\cK_0 \equiv Q\cK$
and observe
\begin{equation}
\tau(a) = \tau(\jed \cdot a) = \frac{1}{2} \tau( \{\jed, a \}) = \frac{1}{2} \{ \tau(\jed), \tau(a) \}
= \frac{1}{2} Q\tau(a) + \frac{1}{2} \tau(a) Q.
\end{equation}

Hence $\tau(a) Q = Q \tau(a)Q$ and $Q\tau(a) = Q \tau(a) Q$. Thus
 $[Q, \tau(a)] = 0$
for any $a \in \cB(\cH)$ and one can restrict oneself to unital Jordan morphism, 
which will be also denoted by $\tau$. To see this, we note that, 
for any $f \in \cH$ and $a \in \cB(\cH)$ one has
\begin{eqnarray}
\label{Jordan1}
\lefteqn{\alpha(a)f = W^*\tau(a) W f = \alpha(a \cdot \jed)f = W^* \tau(a \cdot \jed)Wf
= \frac{1}{2} W^* \tau(\{a,\jed \}) Wf}
                                               \nonumber\\
& & {} = \frac{1}{2} W^*\{ \tau(a), Q\} Wf = W^* Q \tau(a)QWf = W_Q^* \tau(a) W_Q f,
\end{eqnarray}

where $W_Q \equiv QW$ is an isometry.
Consequently, it is enough to consider $\tau : \cB(\cH) \to \cB(\cK_0)$, 
$W: \cH \to \cK_0$ such that $W^*W = \jed$ (so for simplicity of notation we drop the subscript $Q$ and 
put $Q = \jed$).

Having well described the facial structure of the set of all states (see Sections 2 and 3) we wish to examine the 
regularity of decomposable map, in the Schroedinger picture, with respect to this structure. 
To this end, we
denote by $\varrho \in \cB(\cH)$ the density matrix determining a state $\phi \in \cB(\cH)^*$. One has
\begin{equation}
U_W: \cB(\cK_0) \to \cB(\cH); \quad
U^*_W: \cB(\cH)^* \to \cB(\cK_0)^*
\end{equation}
and
\begin{equation}
(\phi \circ U_W)(b) = Tr_{\cH}\Bigl( \varrho W^* b W \Bigr) = Tr_{\cK_0}\Bigl( W \varrho W^* b \Bigr) 
= (U^*_W \phi)(b)
\end{equation}
where $b \in \cB(\cK_0)$.

Let $\varrho \in F$ where $F \subset \cS(\cB(\cH))$ is a face.
We recall (see Section 2) that each face in the set of all states (for the finite dimensional case 
each state is a normal one) is a projective face, i.e.
there exists an (orthogonal) projector $p$ such that (see Section 2 or \cite{A2} for a 
recent account of the theory):
\begin{equation}
\label{6.42}
 F \equiv F_p = \{ \phi \in \cS(\cB(\cH)); \quad \phi(p) =1. \}
\end{equation}

On the other hand, let $p$ be a one dimensional projector, i.e. 
$p = |f><f| \equiv p_f$ for some $f \in \cH$.
We observe
\begin{equation}
WpW^* = |Wf><Wf| \equiv |\xi><\xi| \equiv p_{\xi}
\end{equation}
and
\begin{equation}
||Wf||^2 = (Wf, Wf) = (W^*Wf,f) = ||f||^2
\end{equation}
Consequently, $U^*_W(p_f)$ is a pure state.

Now, let us turn to an analysis of Jordan morphism. We begin with the Heisenberg picture
and recall (see \cite{BR}, \cite{KadRin}) for a Jordan morphism $\tau:\cB(\cH) \to \cB(\cK_0)$
there exists a central projection $z \in \tau(\cB(\cH))^{\prime} \cap \tau(\cB(\cH))^{''}$
such that
$$ a \mapsto \tau(a)z  $$
is a morphism, and
$$ a \mapsto \tau(a)(\jed - z) $$
is an antimorphism.

Hence, passing to the Schroedinger picture, for a pure state $\varrho = p_f \equiv |f><f|$, $f \in \cH$,  
and any $a \in \cB(\cH)$ one has
\begin{eqnarray}
\label{Jordan2}
\lefteqn{Tr\{ p_f \alpha(a)\} = Tr \{ p_f U_W \circ \tau(a)\} = Tr \{ W p_f W^* \tau(a) z \}
+ Tr\{ Wp_f W^* \tau(a) (\jed - z) \}   }
                                               \nonumber\\
& & {} = Tr \{ z p_{\xi} z \tau(a)\} + Tr\{ (\jed - z) p_{\xi} (\jed - z) \tau(a)\} 
= \lambda Tr\{ \tau^*_1(p_{\xi_1}) a \}
+ \lambda^{\prime} Tr \{ \tau^*_2 (p_{\xi_2}) a \}
\end{eqnarray}

where $\xi_1 \equiv ||z \xi||^{-1} z \xi$, $\xi_2 \equiv ||(\jed - z) \xi||^{-1} (\jed - z) \xi$,
$\tau_1(\cdot) \equiv \tau(\cdot) z$, $\tau_2(\cdot) \equiv \tau(\cdot)(\jed - z)$, 
$\lambda \equiv || z \xi||^2$, and $\lambda^{\prime} \equiv || (\jed - z) \xi||^2$.
We note that the Pythagorean theorem implies 
\begin{equation}
\label{Jordan3}
\lambda + \lambda^{\prime} = 1.
\end{equation}

Having this explicit form of decomposable maps we can examine 
regularity properties of these maps with respect to the facial structure (cf Section 2). In particular,  
we are in position to compare 2D and 3D cases and to explain the striking 
difference between these two cases.
We begin with 2D case for which there is only one non-trivial family of projectors: one dimensional ones.
Moreover, each non-trivial face (in the set of all states) is determined by 
such projector (cf. \ref{6.42}). 
On the other hand, \ref{Jordan2} and \ref{Jordan3} imply that 
in the Schroedinger picture any genuine decomposable map, i.e.,
neither plain morphism nor plain antimorphism, sends a non-trivial 
face (one dimensional projector) to the convex
combination of two projectors. We know (cf. Section 2) that the smallest face containing two 
(one dimensional) projectors
is a 3-ball. But a 3-ball for 2D case is just the set of all states! Therefore, there is no room 
for other maps and we arrive to the geometrical explanation why any 
positive map, in 2D case, is decomposable.

Turning to 3D-case, our first observation is that this case is equiped with much ``richer'' geometry.
Namely, there are two non-trivial families of projectors: one and two dimensional ones. 
Therefore, the facial structure of 3D system is richer. A repetition of the argument based on 
\ref{Jordan2} and \ref{Jordan3} says that in the Schroedinger picture, the non-trivial decomposable map 
sends one-dimensional projective faces into the 3-balls, which is not the set of all states for this case!
Similar analysis performed, now, for projective faces determined by two dimensional projections 
shows global invariance of the family of projective faces (what is expected! cf Theorem \ref{ASJor}).
Therefore, a certain ingredients of the facial structure are preserved what is not expected for any plain 
positive map. Consequently, there is room for more general maps than decomposable ones.
This explains, from the geomerical point of view why non-decomposable maps can appear in 3D-case.



\section{POSITIVE MAPS VERSUS ENTANGLEMENT }

Positive maps as well as quantum correlations exhibit their non-trivial features only when they are 
defined on non-commutative structures, so in the quantum mechanics setting. Hence,
it is not surprising that  
the concept of entanglement, strictly related to quantum correlations (see \cite{Ma2}) plays an 
important role in quantum computing \cite{Albert}, \cite{Keyl}. 
Its analysis indicates that there is a need for an operational measure of entanglement. 
This demand is strenghtened by the observation that the number of states that can be used for
quantum information is measured by the entanglement.
On the other hand, the programme of classification of entanglement (so quantum correlations) seems to be 
a very difficult task.
In particular, it was realized that the first step must presumably take
the full classification of all positive maps, see \cite{H};  
as a consequence this fact has revitalized the theory of 
positive maps in Physics. This topic has always been studied in Mathematics as can be seen
from the literature (e.g. see  \cite{Choi}-\cite{Choi3}, \cite{K}, \cite{K1}-\cite{Kyecan},
\cite{SW}-\cite{Wor}).

\smallskip

To see the relation between positive maps and entanglement, from a physical point of view, let us  take
a positive map $\alpha_{1,t}: \cA_1 \to \cA_1$, ($\cA_1 \equiv \cB(\cH_1)$), 
$t$ being identified as a time parameter, and consider
the evolution of a density matrix $\varrho$ 
(where $\varrho$ determines the state in $\cS(\cA_1 \otimes \cA_2)$). 
In other words,  we wish to study  $(\alpha_{1,t} \otimes id_2)^d\varrho$.
Here $(\alpha_{1,t} \otimes id_2)^d$ stands for the dual map, i.e. for the dynamical map in
the Schr\"odinger picture. 
Then, if $\varrho$ is an entangled state,
$(\alpha_{1,t} \otimes id_2)^d\varrho$ may develop negative eigenvalues and thus loose consistency as 
a physical state.
That observation was the origin of rediscovery, now in a physical context, of Stinespring's argument
saying that the tensor product of transposition with the identity map
can distinguish various cones in the tensor product structure (see \cite{Sti}, \cite{W}).
This led to the criterion of separability (\cite{P}, \cite{H}) saying that only separable states
are globally invariant with respect to the familly of all positive maps.

\smallskip
It is known (\cite{Wor}, \cite{Choi}) that for the case $M_k(\bC) \to M_l(\bC)$
with $k=2=l$ and $k=2$, $l=3$ all positive maps are decomposable.
A new argument clarifying this phenomenon was presented in Section 6.
Here we note only that for this low dimensional case 
the criterion for separability simplifies significantly. Namely,
to verify separability of a state $\phi$ it is enough to analyse $(\tau \otimes id)^d \phi$, 
with $\tau$ being the transposition, as
other positive maps are just convex combinations of $\C\cP$ maps (they always 
map states into states) and the composition of $\C\cP$ maps with $\tau \otimes id$. This observation is the
essence of the Peres-Horodecki criterion.
We want to add that the lack of normalization for the summands of 
decomposition of a positive map (see Section 6)
does not affect this criterion.

The situation changes dramatically, as we have seen in Section 6,  when both $k$ and $l$ are larger than 2. 
In that case there are plenty of non-decomposable maps 
(see \cite{Kos} and the references given there as well as see 
the preceding Section)  and to analyse entanglement
one cannot restrict oneself to study $\tau \otimes id$.
Thus, a full description of positive maps is needed.
In particular, one wishes to have a canonical form of non-decomposable maps. 
We note that in Section 6 we obtained only some 
clarification of the
nature of decomposable maps. The importance of the former follows from the observation
saying that this class of maps does not contain transposition.
On the other hand, the theory of non-decomposable maps
offers a nice construction of examples of entangled states (see \cite{HKP}). However, 
the classification of non-decomposable
maps is a difficult task which is still not completed (\cite{St3}, \cite{LMM}).
Nevertheless, it seems that such classification is an indispensable step for an operational 
generalization of the Peres-Horodecki criterion.

We want to close the section with another important remark concerning the relation
between quantum correlations and entanglement. Following the idea of coefficients of independence from 
classical probability calculus and working within the framework of non-commutative integration theory
one can define (see \cite{Ma2} and \cite{Ma3}) the coefficient
of quantum correlations. 
If the coefficient 
of quantum correlations is equal to zero
for any $A \in \cA_1 \otimes \cA_2$ then, using the description of locally
decomposable maps, we proved that the state $\phi$ is separable.
These observations provide the complementary approach to entanglement and the just quoted
result shows how strong is the interplay between separability and certain subtle
features of positive maps. However, this is not unexpected as the indicated correspondence
between Schr\"odinger's and Heisenberg's picture relies on the underlying algebraic structure and
geometry of the state space, see Sections 1 and 2 as well as \cite{A}, \cite{C}, and \cite{E}.
Nevertheless, it should be stressed that the complete description of quantum correlations
as well as the full classification of all positive maps are still open and challenging problems.

\section{Appendix: Glossary}
In order to make the paper more accessible to readers not really familiar with 
abstract mathematical terminology we add a glossary, in which the basic notions are defined
and some basic facts are noted. 
The theory of \Cs-algebras can be found in the books of
\cite{BR}, \cite{KadRin} while the geometry of states is described in  \cite{A} and \cite{A2}.
\begin{itemize}
\item{A1.} Let $K$ ($K^{\prime}$) denote a convex set of a real vector space $X$ ($X^{\prime}$ respectively).
A map $\alpha : K \to K^{\prime}$ is called affine if the following property holds:
$\alpha(\lambda k_1 + (1 - \lambda) k_2) = \lambda \alpha(k_1) + (1 - \lambda) \alpha(k_2)$
for all $k_1, k_2 \in K$ and $0 \le \lambda \le 1$.
\item{A2.} A Jordan algebra over $\bR$ is a real vector space $\cA$ equippped with a commutative 
bilinear product $\circ$ that satisfies the identity
$$(a^2 \circ b) \circ a = a^2 \circ (b \circ a)$$
for all $a,b  \in \cA$.
\item{A3.} An associative algebra $\cA$ (linear space equipped with associative multiplication) with 
involution $^*$ is called $^*$-algebra. When on $\cA$ is defined a norm and $\cA$ is complete with respect
to this norm, $\cA$ is called a Banach $^*$-algebra. Finally,
a Banach $^*$-algebra $\cA$ is called a \Cs-algebra
if it satisfies $||a^*a|| = ||a||^2$ for $a \in \cA$.
\item{A4.} For a subset $Y$ of $\cB(\cH)$ the set of operators in $\cB(\cH)$ that commute with all operators
in $Y$ is called the commutant of $Y$ and is denoted by $Y^{\prime}$. 
\item{A5.} A von Neumann algebra on $\cH$ is a $^*$-algebra $\cM$ of $\cB(\cH)$ such that $\cM = \cM^{"}$
( $^"$ stands for the double commutant). Another name for an (abstract) von Neumann algebra is 
$W^*$-algebra.
\item{A6.} A homomorphism between two \Cs-algebras, $\cA_1$ and $\cA_2$,
is a map $\Phi: \cA_1 \to \cA_2$ preserving
the algebraic structures, i.e. $\Phi$ is linear, $\Phi(ab)= \Phi(a)\Phi(b)$ and $\Phi(a^*) = \Phi(a)^*$.
If an inverse homomorphism exists, $\Phi$ is called isomorphism.
\item{A7.} A state on a \Cs-algebra $\cA$ is a linear functional $\omega: \cA \to \bC$
which is positive (i.e. $a \ge 0$ implies $\omega(a) \ge 0$) and normalized (i.e. $\omega(\jed)=1$). 
The set of all states of $\cA$ will be called the state space and denoted by $\cS$. The state space $\cS$ of $\cA$ is a ($w^*$-compact)
face of the unit ball of $\cA^*$.
\item{A8.} Normal states $\omega$ on $\cB(\cH)$ are of the form $\omega(a) = Tr (\varrho\, a)$ where $\varrho$
is a uniquely determined positive operator on $\cH$ having the trace $Tr$ equal to $1$.
The set of all normal states on $\cB(\cH)$ will be called the normal state space. There is no useful compact topology on the 
normal state space of $\cB(\cH)$ for a general $\cH$. However, the normal state space of a von
Neumann algebra $\cM$ is a face of the state space $\cS$ of $\cM$.
\item{A9.} {\it GNS-representation}: if $\cA$ is a \Cs-algebra and $\omega$ is a state on $\cA$, then there exists a Hilbert space
$\cH_{\omega}$, a cyclic unit vector $\Omega \in  \cH_{\omega}$ and a representation
$\pi_{\omega}$ of $\cA$ on $\cH_{\omega}$ such that $\omega(a) = (\Omega, \pi_{\omega}(a) \Omega)$.
\item{A10.} A face $F$ of the state space $\cS$ of \Cs-algebra $\cA$ is exposed 
iff there exists a $a \in \cA$ and an $\alpha \in \bR$ such that 
$x(a) = \alpha$ for all $x \in F$ and $x(a) > \alpha$ for all $x \in \cS \setminus F$.
\item{A11.} A face $F$ of the normal state space $K$ of $\cB(\cH)$ is said to be norm exposed if there exists an 
$a \in \cB(\cH)$, positive, such that $F = \{ \sigma \in K; \sigma(a) =0 \}$. 
A norm closed face $F$ of the normal state space of $\cB(\cH)$ is norm exposed.
In general, a face $F$ of a compact subset $K$ in a vector space $V$ is norm exposed 
if there is a positive bounded affine functional
$\phi$ on $K$ whose zero set equals $F$.
\item{A12.} An element $p \in \cA$ is called a projector if $p^* =p$ and $p = p^2$.
\item{A13.} Two projections $e$ and $f$ in $\cB(\cH)$ are said to be equivalent if there exists
$v \in \cB(\cH)$ such that $v^*v = e$ and $v v^* = f$.
\item{A14.} The convex hull of a subset $E$ of a real vector space $X$ consists of all elements of the form 
$\sum_{i=1}^n \lambda_i x_i$ where $x_i \in E$, $\lambda_i \ge 0$ for $i=1,...,n$ and 
$\sum_{i=1}^n \lambda_i = 1$. It will be denoted by $co(E)$.
\item{A15.} The $\sigma$-convex hull of a bounded set $F$ of elements in a Banach space is the set of all
sums $\sum_i \lambda_i x_i$ where $\lambda_1,...$ are positive scalars with sum 1 and $x_1,...$ are 
elements of $F$.
\item{A16.} An ordered normed vector space $V$ with a generating cone $V^+$ is said to be a base norm space if 
$V^+$ has a base $K$ located on a hyperplane $H$ ($ 0 \not\in H$) such that the closed unit ball of 
$V$ is $co(K  \cup - K)$. The convex set $K$ is called the (distinguished) base of $V$.
An order unit space is an an ordered normed vector space $V$ over $\bR$ with a closed positive cone 
and an element $e$, satisfying
$$||a|| = inf \{ \lambda>0; - \lambda e \le a \le \lambda e \}$$
for any $a \in V$.
\item{A17.} A lattice is a set with an order relation such that 
every pair of elements $p,q$ has a least upper bound 
(denoted by 
$p \vee q$) and a greatest lower bound (denoted by $p \wedge q$).
A lattice $L$ is complete if every subset has a least upper bound.
\item{A18.} Let $\cF$ stand for the set of projective faces of the normal state space of $\cB(\cH)$,
ordered by inclusion. Define the map $F \mapsto F^{\prime}$ on $\cF$ by $(F_p)^{\prime} 
= F_{p^{\prime}}$ for each projector $p$ ($p^{\prime} \equiv 1 - p$). $F^{\prime}$ is called the 
complementary face of $F$.
\item{A19.} Let $\cM \subset \cB(\cH)$ be a $^*$-algebra. The set 
$ \cZ \equiv \cM^{\prime} \cap \cM^{''}$
is called the centrum of the von Neumann algebra $\cM^{''}$. 
A projector $p \in \cZ$ is called a central projector.
\item{A20.} Let $Y$ be a subspace of a vector space $X$. The equivalence class (coset) of an element
$x \in X$ with respect to $Y$ is denoted by $x + Y$ and is defined to be the set $x + Y = 
\{v; v = x + y, y \in Y \}$.
The name {\it equivalence} stems from the definition saying that two elements $x,z$ of $X$ are 
$Y$-equivalent whenever $x - z \in Y$.
It can be shown that under algebraic operations defined by $(w + Y) + (x + Y) = (w + x) + Y$
and $\lambda (x + Y ) = \lambda x + Y $, $\lambda \in \bC$, these classes constitute the 
elements of vector space. This space is called the quotient space and it is denoted by $X/Y$.
\end{itemize}

\section{Acknowledgements}
The author would like to thank Marcin Marciniak for many fruitful discussions. Thanks also are
due to Fabio Benatti, Stanislaw Kryszewski, L. L. Labuschagne and Boguslaw Zegarlinski 
for their careful reading of the manuscript.
The support of grant PBZ-MIN-008/PO3/2003 and
Poland-South Africa Cooperation Joint project 
is gratefully acknowledged.

\end{document}